\documentclass[useAMS,usenatbib]{mn2e}
\usepackage{graphicx}
\begin{document}
\newcommand{\comment}[1]{{\ }}
\newcommand{\degree}{^{\circ}}
\def\Msun{\ifmmode{~{\rm M}_\odot}\else${\rm M}_\odot$~\fi}
\newcommand{\der}[2]{{\frac{d#1}{d#2}}}
\newcommand{\doo}[2]{{\frac{\partial#1}{\partial#2}}}
\newcommand{\fat}[1]{{\bmath{ #1}}}
\newcommand{\FAT}[1]{{\mathbfss{ #1}}}
\title[Secular motion under quadratic perturbation]
{Computing secular motion under slowly rotating quadratic perturbation}

\author[S. Mikkola and P. Nurmi]
{
Seppo Mikkola\thanks{E-mail:seppo.mikkola@utu.fi} and Pasi Nurmi\thanks{E-mail:pasi.nurmi@utu.fi}
\\
Tuorla Observatory, University of Turku, V\"ais\"al\"antie 20, Piikki\"o, Finland
}

\date{\today}

\pagerange{\pageref{firstpage}--\pageref{lastpage}} \pubyear{}

\maketitle

\label{firstpage}

\begin{abstract}
We consider secular perturbations of nearly Keplerian two-body
motion under a perturbing potential that 
can be approximated to sufficient accuracy by expanding it to second
order in the coordinates. After averaging over time to obtain the secular
Hamiltonian, we use angular momentum and eccentricity vectors as elements.
The method of variation of constants then
leads to a set of equations of motion  that 
are simple and regular, thus allowing
efficient numerical integration.  Some possible applications are briefly
described.
\end{abstract}

\section{Introduction}
If a binary star, or a planet revolving around a star, is perturbed
by one or more distant relatively slowly moving masses the
perturbation can be approximated by a tidal field. Fast computation
of secular perturbations in such a potential is thus desirable.
Potential applications include the Kozai resonance, motion of a
well isolated binary in a star cluster and cometary motion 
under the galactic tidal field.

Several authors have considered the computation of cometary motion
perturbations due to the galactic tidal field e.g. 
\cite{1986Icar...65...13H}, 
\cite{1999Icar..137...84W}, 
\cite{2001MNRAS.324.1109B}. 
More recently \cite{2004MNRAS.349..347F} 
suggested the use of the time averaged secular perturbing function
and the Lagrangian equations for computing the evolution of orbital elements
of a comet.
While 
\cite{2005CeMDA..93..229F} 
compared various ways to compute the evolution of orbital elements (using Keplerian
or Delaunay elements).

\cite{2005MNRAS.364.1222B} 
used the so called vectorial elements, i.e. the areal velocity vector and the
eccentricity vector. However, the perturbing potential
considered in all these studies was a simplified one, in most cases
just the term causing the force perpendicular to the galactic plane 
was included. 

In this paper we consider a general quadratic potential. For more
mathematical generality we include the possibility of a constant 
force (although such a term does not appear in a tidal field) and
also a slow rotation of the quadratic perturbing field is allowed.
We form the simple regular equations of motion using the vectorial
elements. 

Finally it is necessary to stress that the main purpose 
of this paper is to introduce the simple secular equations of
motion when a two-body system is perturbed by a (at most) quadratic
perturbing potential.

\section{Perturbations of vectorial elements}

\subsection{Perturbing function}
Consider a system in which the Keplerian motion of a particle
around a central mass $m$ is perturbed by a perturbing potential 
$U(\fat r,t)$. In addition, a slow rotation (with angular velocity $\fat N$)
is assumed. The Hamiltonian may be written 
\begin{equation}
H=\frac12 \fat v^2-\frac{m}{|\fat r|}-U(\fat r, t)=-\frac{m}{2a}-U(\fat r, t),
\end{equation}
where $m$ is the mass of the central body, $\fat r$ is the position vector
with components $(x_1,x_2,x_3)$, 
$\fat v$ is the velocity vector with components $(v_1,v_2,v_3)$ 
i.e. $\dot \fat r=\fat v$ (which
also can be identified as the momentum vector) and $a$ is the semi-major axis of
the Kepler motion. The function $U$
is the perturbing function, assumed at most quadratic in the coordinates.  
If the time dependence of $U$ is due to rotation of the potential 
with a constant angular velocity, then one can
consider the system in the rotating coordinate system in which 
the perturbing function is a constant. Thus, using the suffix notation, 
one may write in place of $U$
\begin{equation}\label{perturbingfunction}
\widetilde U=  f_ix_i+\frac12 x_i G_{ij} x_j+ \epsilon_{ijk}N_ix_jv_k
,\end{equation}
where there is no time dependence, $f_i$ are components of a (possible)
constant acceleration vector $\fat f$ and $ G_{ij}$ is the second derivative matrix: 
\begin{equation} f_i={\doo{\widetilde U}{x_i}}_{|\fat r=\fat 0};\ \ \ \ 
G_{ij}={\frac{\partial^2\widetilde U}{\partial x_i\partial x_j}}_{|\fat r=\fat 0}.\end{equation}
The last term in (\ref{perturbingfunction}) is the scalar product of the
angular velocity vector $ \fat N$ of the perturbing
potential and the angular momentum vector of the moving particle, 
here written using the Levi-Civita symbol $\epsilon_{ijk}$.

Since the matrix $ G_{ij}$ is symmetric it can be diagonalized by selecting
a suitable coordinate system. However, there may be situations in which this is not 
convenient. Therefore we consider the general non-diagonal case.
In the theory presented here that does not add any substantial complications.

\subsection{Averaging the Hamiltonian}

The auxiliary quantities needed in averaging the Hamiltonian are:
\begin{enumerate}

\item
The angular momentum vector per unit of mass (actually areal speed)
 of the particle
\begin{equation}
{\fat J}={\fat r\times \fat v}.
\end{equation}

\item
The eccentricity vector
(also known as the Runge-Lenz- or Laplace vector)
\begin{equation}
{\fat e}=\frac{\fat v\times \fat J}{m}-\frac{\fat r}{r},
\end{equation}
or, for our purposes, the vector
\begin{equation}
{\fat E}=\sqrt{ma}\fat e
\end{equation}
is more convenient.
\item
The mean values of the position vector components 
\begin{equation}
<{x_i }>=-\frac32 a  e_i=-\frac{3}{2}\sqrt{\frac{a}{m}}\fat E_i.
\end{equation}

\item
The time averaged coordinate products can be written
\begin{equation}
<x_ix_j>=\frac{a}{2m} {\fat J}^2\delta_{ij}-\frac{a}{2m} J_iJ_j+\frac{5a}{2m} E_i E_j,
\end{equation}
where $\delta_{ij}$ is the Kronecker $\delta$. This result was easy to derive
for $<x_3^2>$. The generalization was then formed by educated guess and
use of computer algebra for confirmation.
\end{enumerate}
The secular perturbing function $R=<\widetilde U>$ takes the form
\begin{eqnarray}
R=f_i<x_i>+      
\frac{a}{4m}
 G_{ij}<x_ix_j>
+ N_i J_i
.\end{eqnarray}

If one  introduces the matrixes $ A_{ij}$ and $ B_{ij}$ with the elements
\begin{eqnarray}
A_{ij}&=&\frac{a}{2m} \left((\textstyle{\sum_kG_{kk}})\delta_{ij}-G_{ij}\right)\\
B_{ij}&=&\frac{5a}{2m} G_{ij},
\end{eqnarray}
and the vector $\fat F$
\begin{equation}
\fat F=-\frac32 \sqrt{\frac{a}{m}}\fat f,
\end{equation}
then the averaged perturbing function can be written
in the simple form
\begin{equation}\label{quadraticH}
R=F_i E_i+\frac12  J_i A_{ij} J_j+\frac12  E_i B_{ij} E_j+ N_i J_i,
\end{equation}
where the vectors  $\fat F$, \  $\fat N$, and matrices  
$\FAT{A}$, $\FAT{B}$ are constants in any 
particular orbit.

\subsection{Equations of motion}
The reason for the choice of the elements $\fat J$ and $\fat E$ was that
one obtains for their components the simple Poisson bracket relations
\begin{eqnarray}\textstyle{
\{J_i,J_j\}=\epsilon_{ijk}J_k, \: \{E_i,E_j\}=\epsilon_{ijk}J_k,\: \{J_i,E_j\}=\epsilon_{ijk}E_k.
}
\end{eqnarray}
With these formulae one gets the 
equations of motion (Breiter and Ratajczak(2005) and references therein)
\begin{eqnarray}\label{Jequation}
\dot {\fat J}&=-\{{\fat J},R\}=&{\fat J\times}\doo{R}{\fat J}+{\fat E\times}\doo{R}{\fat E} \\
\dot {\fat E}&=-\{{\fat E},R\}=&{\fat E\times}\doo{R}{\fat J}+{\fat J\times}\doo{R}{\fat E},
\label{Eequation}
\end{eqnarray}
and by (\ref{quadraticH})
the partial derivatives of $R$ with respect to the vectorial elements are
\begin{eqnarray}
\doo{R}{\fat J}&=&\fat N+\FAT{A}\fat J\\
\doo{R}{\fat E}&=&\fat F+\FAT{B}\fat E,
\end{eqnarray}
where we have used the vector-matrix notation.
In deriving the above equations one should note that 
$\fat J$ and $\fat E$ are constants of motion
under the two-body Hamiltonian $H=-m/(2a)$. Thus 
$a$ can be considered a (numerical) constant since $\{{\fat J},a\}=\{{\fat E},a\}=\fat 0$.
The equations of motion have the integrals
\begin{equation}
{\fat J \cdot\fat E =0},\ \ {\fat J}^2+{\fat E}^2=ma,\ \ R={\rm constant}.
\end{equation}
Here we must remark that the form of the perturbing function is not unique.
This is because of the larger-than-necessary number of variables and 
because one may e.g add terms proportional to ${\fat J}^2+{\fat E}^2$ and/or
to $\fat J \cdot \fat E $ which would not affect the final expressions
for the derivatives of the elements in (\ref{Jequation}) and (\ref{Eequation}). 

Note that when the rotation term $ {\fat N\cdot\fat J}$ is included, the result for the evolution
of the vectorial elements will be obtained in the rotating coordinate system.

\section{Applications}

Possible applications of the vectorial element equations include:
\begin{enumerate}
\item
Constant force, or rotating constant force. 
This may be restricted to toy applications, but
was included in our treatment for the sake of generality and because
it is mathematically simple.

\item
Isolated binary in a star cluster. Assuming the perturbing stars
are distant and move relatively slowly, one may approximate
the perturbing potential by
\begin{equation}
U=\sum_k\frac{m_k}{2|\fat R_k|^3}\left( r^2-3\frac{(\fat R_k\cdot \fat r)^2}{|\fat R_k|^2} \right),
\end{equation}
where $\fat R_k$ are the positions of the perturbers relative 
to the centre-of-mass of the binary.
Writing $\fat R_k=(X_{k1},X_{k2} ,X_{k3})$ we get for the components
of the matrix $\FAT{G}$  
\begin{equation}
G_{ij}=\sum_k\frac{m_k}{|\fat R_k|^3}\left(\delta_{ij}-3 X_{ki} X_{kj} \right),
\end{equation}
where the matrix elements may be time dependent, but this does not invalidate
the equations presented above.

\item
Kozai resonance. This term is often used in connection of triple systems
in which averaging over the outer orbit gives the perturbing potential of the form
\begin{equation}U=\frac{m_3}{8b_3^3}(r^2-3(\fat z\cdot \fat r)^2),\end{equation}
where $m_3$ is the mass of the third body, $b_3$ is the semi-minor axis of the
outer ellipse and $\fat z$ is the unit normal vector of the orbital plane 
of the distant body.  In this case we have
\begin{equation}
G_{ij}=\frac{m_3}{4b_3^3}(\delta_{ij}-3z_iz_j),
\end{equation}
where $z_k$ are the components of the vector $\fat z$.

\item
Cometary motion under the galactic tide. In this case
the tidal potential is usually written
\begin{equation}\widetilde U=\frac12\sum_k G_{kk}x_k^2+\fat N\cdot\fat J,\end{equation}
which is possible by choosing the coordinate system such that $x_3$ axis
is perpendicular to the galactic plane and the $x_1$-axis points to the galactic centre.

\item
The perturbing potential due to a small disk of mass can also be expressed 
easily in terms of the vectorial elements, although not in quadratic form.
The secular perturbing function for this case is often written in terms of the
Delaunay elements in the form
$
R=\epsilon L^{-3}G^{-3}(1-3G^{-2}H^2),
$
where $\epsilon$ is a (small) constant and $L,\ G,\ H$ are the Delaunay elements.
Here $L=\sqrt{m a}$ is a constant,  $G=|\fat J|$ and  $H=\fat z\cdot\fat J$,
where $\fat z$ is the unit direction vector of the symmetry axis of the potential.
Thus 
\begin{equation}
R=\epsilon L^{-3}|\fat J|^{-3}(1-3\frac{(\fat z\cdot \fat J)^2}{|\fat J|^2})
.\end{equation}
Consequently it is easy to include the effect of any such potential into the 
equations of motion for  the vectorial elements.
Note that this kind of a term could be used to approximate the effect
of the planets (averaged over orbits)
to comets that come close, but not too close, to the Solar System.
\end{enumerate}

\section{Numerical aspects}
We ran some numerical
test computations  and compared results against direct numerical
integration of the equations of motion for the coordinates.
The tests used the tidal field of the Galaxy
in the form $\widetilde U=\frac12\sum_k G_{kk}x_k^2+\fat N\cdot \fat J$ (parameters from \cite{2005CeMDA..93..229F}) 
and a value for the
semi-major-axis of an Oort Cloud comet of $a\ge 10,000$AU. 

\bigskip
\noindent
The experiments suggest:

If one needs results for a long time and infrequent output is allowed,
then one of the best numerical integrators available is the Burlirsh-Stoer
method (actually the code known as $\scriptstyle{\rm \bf DIFSY1}$, originally written by Burlirsh).
However, this is efficient only when run with optimally long
steps, which may be too long for some applications.

If one needs frequent output, then
the method of choice is the implicit midpoint method. It is very simple to
implement and gives quite high precision, especially it conserves the
quadratic integrals of motion (\cite{HuangLeimkuhler}, \cite{SanzSernaCalvo}).

In this case one finds that the stepsize equal to one period (i.e. $10^6$ years)
gives a clearly satisfactory accuracy: the plots of the vectorial
elements from both computations (secular and coordinate integration) 
agree well, except for a small phase error.

However, for $a>> 20,000$AU the results differ 
and the reliability of a secular theory becomes questionable.

\section{Conclusions}

We have generalized the equations of \cite{2005MNRAS.364.1222B} who 
considered the simple case of perturbing function of the form $g_{33} z^2$.

The equations for vectorial elements, 
even in the case of a general quadratic potential
are simple and regular, 
contrary to what one obtains when using Keplerian or Delaunay elements
(\cite{2005CeMDA..93..229F}).

The integration of the secular theory equations is typically
faster than direct coordinate integration  by  two orders of magnitude.

The implicit midpoint method may be even faster than  Burlirsh-Stoer,  
but there is no simple way of knowing accuracy and optimal stepsize.

The most reliable integration method is Burlirsh-Stoer extrapolation.

For comets the secular theory results are not accurate for $a>>20,000$AU,  
as comparison with direct integration shows (simply: perturbation can be too large).

For an integration over only one  period at a time the midpoint method and
Burlirsh-Stoer are nearly equally fast. Thus in this case the midpoint method
may be preferable due to its simplicity.

\label{lastpage}
\end{document}